\documentclass[preprint,12pt]{elsarticle}
\usepackage{mathtools}
\usepackage[most]{tcolorbox}
\usepackage{underscore}
\usepackage{threeparttable} 
\usepackage{booktabs}
\usepackage{multirow}
\usepackage{makecell}
\usepackage{listings}
\usepackage{csquotes}
\usepackage{url}
\usepackage{adjustbox}
\usepackage{fontawesome}
\usepackage{url}

\newcolumntype{C}[1]{>{\centering\let\newline\\\arraybackslash\hspace{0pt}}m{#1}}

\renewcommand*\thelstnumber{\makebox[3em][r]{\ifnum\value{lstnumber}<10 0\fi\the\value{lstnumber}}}
\AtBeginDocument{%
\newtcblisting[blend into=listings]{ccode}[1][]{%
  colframe=black,
  enhanced,
  listing only,
  listing remove caption=false,
  listing options={  
    basicstyle=\scriptsize\ttfamily, 
    showstringspaces=false, 
    breaklines=true,
    breakatwhitespace=false,
    extendedchars=true, 
  },
  size=fbox,
  coltitle=black,
  attach boxed title to bottom center={yshift=-10pt},
  boxed title style={enhanced jigsaw, colback=white, sharp corners, boxrule=0pt},
  #1
}
}

\lstdefinelanguage{myC++}{
    breaklines=true,
    breakatwhitespace=true,
    language = c++
}

\lstdefinelanguage{myC++}{
    breaklines=true,
    breakatwhitespace=true,
    language = c++
}

\tcbset{
    snippetraster/.style={
        enhanced,
        raster columns=2,
        raster equal height=rows,
        raster after skip=0pt,
        raster left skip=0pt,
        raster right skip=0pt,
        raster column skip=0pt,
        fonttitle=\bfseries,
        listing only,
        listing options={
            aboveskip=2pt,
            belowskip=0pt,
            basicstyle=\ttfamily\scriptsize, 
            language=myC++  
        },
        raster column 1/.style={
            sharp corners=east,
            before title={},
            title style={
                font=\bfseries\footnotesize,
                colframe=white, 
                colback=none, 
                overlay={\draw (frame.south east) -- (frame.north east);}, 
                interior hidden 
            },
            rightrule=0pt,
            size=small,
            underlay={
                \node[minimum width=1cm,minimum height=0.5cm,outer sep=auto,
                anchor=north east,fill=white] at (interior.north east)
                {\itshape\small Request};
        }
        },
        raster column 2/.style={
            colback=black!20,
            sharp corners=west,
            before title={},
            title style={
                font=\bfseries\footnotesize,
                colframe=white, 
                colback=none, 
                overlay={\draw (frame.south west) -- (frame.north west);}, 
                interior hidden 
            },
            leftrule=0pt,
            size=small,
            underlay={
                \node[minimum width=1cm,minimum height=0.5cm,outer sep=auto,
                anchor=north east,fill=white] at (interior.north east)
                {\itshape\small Response};
            }
        },      
    }
}

\usepackage{amssymb}

\usepackage{tabularx}
\journal{Computer \& Security}
\usepackage{xspace}
\definecolor{darkorange}{RGB}{210,105,30}
\newcommand{\DISINFOX}{{\sffamily DISINFOX}\xspace}

\begin{document}
\sloppy

\begin{frontmatter}
 


\title{Toward interoperable representation and sharing of disinformation incidents in cyber threat intelligence}


\author[inst1]{Felipe Sánchez González}
\author[inst2]{Javier Pastor-Galindo}
\author[inst1]{José A. Ruipérez-Valiente}
\affiliation[inst1]{organization={Department of Information and Communications Engineering, University of Murcia},
            postcode={30100}, 
            country={Spain}}

\affiliation[inst2]{organization={Computer Systems Department, Universidad Politecnica de Madrid}, 
             postcode={28031}, 
             country={Spain}}

\begin{abstract}
A key countermeasure in cybersecurity has been the development of standardized computational protocols for modeling and sharing cyber threat intelligence (CTI) between organizations, enabling a shared understanding of threats and coordinated global responses. However, while the cybersecurity domain benefits from mature threat exchange frameworks, there has been little progress in the automatic and interoperable sharing of knowledge about disinformation campaigns. This paper proposes an open-source disinformation threat intelligence framework for sharing interoperable disinformation incidents. This approach relies on i) the modeling of disinformation incidents with the DISARM framework (MITRE ATT\&CK-based TTP modeling of disinformation attacks), ii) a custom mapping to STIX2 standard representation (computational data format), and iii) an exchange architecture (called \DISINFOX) capable of using the proposed mapping with a centralized platform to store and manage disinformation incidents and CTI clients which consume the gathered incidents. The microservice-based implementation validates the framework with more than 100 real-world disinformation incidents modeled, stored, shared, and consumed successfully. To the best of our knowledge, this work is the first academic and technical effort to integrate disinformation threats in the CTI ecosystem.
\end{abstract}




\end{frontmatter}


\section{Introduction}
\label{sec:intro}

While propaganda, deception, disinformation or influence operations are not new phenomena in the geopolitical landscape, they have become increasingly common in recent years due to the reach and scale of modern social networks~\cite{Jethava2024}. Platforms such as X, Facebook, and Instagram have worsened the effects of these practices, as news and opinions from anyone can now go viral with the click of a button~\cite{doi:10.1126/science.ade7138}.

In particular, disinformation has been maligned and used by outsiders to influence public elections, the so-called Foreign Information Manipulation and Interference (FIMI)~\cite{eeasfimi,HybridCoE2022}. As a case in point, the Ukrainian war has provided the perfect scenario for disinformation campaigns to proliferate \cite{ukrfacebookdisinfo}. Social media platforms have been critical in spreading disinformation about war and big political events, influencing users worldwide, and shaping public opinion across nations \cite{ioinsocialmedia}. This use of disinformation has alerted government entities, such as the European External Action Service (EEAS) or the European Centre of Excellence for Countering Hybrid Threats (Hybrid CoE), who recognize the need for structured approaches to model these threats, enabling more informed analysis and understanding of their mechanisms~\cite{eeas1, eeas2, HybridCoE2022, Pamment2020}.

From a cybersecurity perspective, a disinformation campaign could be understood as a set of incidents that deliberately promotes false, misleading, or misattributed information through cyberspace, where social networks are common attack vectors to compromise end-user beliefs~\cite{disinfoiscybersecurity}. Much like cyberattacks, it can erode trust in a company's reputation, destabilize markets by spreading false economic information, or compromise valuable assets by manipulating public perception and decision-making~\cite{10492674}. 

These manipulative efforts are rarely isolated phenomena, as they typically function as part of influence operations, where disinformation is paired with cybersecurity techniques that enhance its effectiveness \cite{Etudo2023}. Therefore, considering that these campaigns are highly dependent on the Internet and social networks, and their nature and characteristics are similar to existing cyber threats, they can be classified as a cybersecurity concern~\cite{Walker2019}. In fact, powerful organizations like the European Union Agency for Cybersecurity (ENISA) state them as one of the main official cyber threats~\cite{enisathreat}.

One of the main cybersecurity countermeasures to gain resiliency and combat risks is Cyber Threat Intelligence (CTI). CTI is a discipline focused on understanding the capabilities, intent, motivations, and opportunities of relevant cyber adversaries and their associated tactics, techniques, and procedures (TTPs) \cite{10117505}. Indeed, traditional CTI solutions such as Cyber Threat Exchanges (CTX) have reached a significant level of maturity and enable the sharing of indicators of compromise (IoCs), CTI reports, and other evidence to increase cybersecurity defenses of CTI consumers~\cite{johnson2016guide}. Platforms such as \textit{AlienVault OTX}, \textit{ThreatFox}, or \textit{DigitalSide Threat-Intel Repository} act as sources of cybersecurity information across thousands of collaborators worldwide, feeding organizations' knowledge of emerging threats by standard methodologies (such as vulnerability assessment like \textit{CVSS} or \textit{CVE}), frameworks (like MITRE ATT\&CK), protocols (such as \textit{TAXII}), and formats (like Structured Threat Information eXpression (STIX2)~\cite{stix2}, a standardized language for representing and sharing cyber threat intelligence). Commonly, CTI end-points like \textit{OpenCTI} or \textit{MISP} locally ingest, fuse, analyze, and correlate the uploaded IoCs and attack techniques to enable a custom proactive reaction~\cite{10146036}.

This successful interoperable effort to counter cyber threats may also be effective against disinformation. This paper explores the possibility of using established CTI and standardized tools and processes to combat and assess disinformation actors, their tools, and objectives~\cite{baraniuk2024potential}. Particularly, the DISARM framework \cite{disarm} aims to map disinformation incidents to TTPs, as is commonly done for cybersecurity incidents in the CTI domain. TTPs are valuable not only for increasing interoperability and integration with CTI platforms but also because they have proven effective in modeling and investigating disinformation incidents over time~\cite{pastor2025influence}. By framing disinformation campaigns within the CTI domain and adapting current CTI tools, it becomes possible to integrate the sharing of disinformation threats through standard formats already understood by CTI solutions and personnel \cite{disinfomodel, HybridCoE2022}.

However, there remains a lack of CTX-specialized repositories where analysts and collaborators can share disinformation incidents in a structured and interoperable way. Such a serving platform would enable the programmatic ingestion and extraction of structured and standardized objects by CTI platforms, facilitating the automatic management of disinformation intelligence. As a result, the use of CTI methodologies to tackle disinformation and influence operations could enhance global collaboration, the detection and mitigation of disinformation campaigns, and provide a unified framework to address these threats by sharing common standards.

To address the identified gaps, this work introduces an open-source, modular, and interoperable disinformation threat intelligence framework designed to model, represent, and share disinformation incidents using CTI methodologies. The development of the framework is structured around the following key objectives:  
\begin{itemize}
    \item Review the existing frameworks proposed for disinformation modeling and select the most appropriate to label disinformation incidents (Section \ref{sec:disinfomodeling}).
    \item Define a mapping between the evidence generated in a disinformation incident and a standardized, computable language such as STIX2, establishing the base data model. Use this mapping to model and structure a real disinformation incident as a use case (Section \ref{sec:mapping}).
    \item Develop an exchange architecture, called \DISINFOX\footnote{\url{https://github.com/CyberDataLab/disinfox}}, as a modular and interoperable client-server platform for sharing disinformation incidents. This includes designing its architecture to effectively leverage the structured data model and validating its full lifecycle, from ingestion and modeling to integration with external CTI platforms such as OpenCTI (Section \ref{sec:design}).
\end{itemize}

The remainder of this paper is as follows. Section \ref{sec:sota} summarizes the state of the art in disinformation modeling, CTX and public disinformation databases. 
Then, Section \ref{sec:disinfomodeling} illustrates the analysis behind choosing a framework to model disinformation incidents.
Later, Section \ref{sec:mapping} describes the translation of a disinformation incident to a standardized and structured language.
Finally, Section \ref{sec:design} describes the main characteristics of the exchange architecture, the nodes that drive the building of the scheme, and the verification with a mature CTI platform. 


\section{State of the Art} \label{sec:sota}

The growing importance of structured and interoperable threat intelligence has led to the rise of several platforms that support different aspects of this process. Among the most widely adopted are EclecticIQ Threat Intelligence Platform~\cite{eclecticiqThreatIntelligence}, MISP~\cite{misp}, OpenCTI~\cite{filigranOpenCTIFiligran}, and AlienVault Open Threat Exchange (OTX)~\cite{alienvaultLevelBlueOpen}. These platforms differ in their approaches, capabilities, and levels of community involvement, yet each plays a vital role in shaping the CTI landscape. 

From the academic perspective, there are recent surveys that analyze current practices and tools regarding CTI \cite{ctiapproaches, ctiminingsurvey}. For example, a platform focused on monitoring and managing cyber threats in the agricultural realm \cite{ctiagriculture} demonstrates the use of CTI in less traditional domains. The Distributed Security Framework for Reliable Threat Intelligence Sharing~\cite{distributedtip} emphasizes the importance of a decentralized approach to enhance the reliability and timeliness of shared threat information. Similarly, the Malware Information Sharing Platform (MISP)~\cite{misp} provides an open-source solution for collecting, storing, and distributing IoC among organizations, promoting collaborative defense mechanisms. Addressing the need for contextual awareness, the Context-Aware Cyber Threat Intelligence Exchange Platform~\cite{contextawaretip} integrates various data sources to enrich the intelligence gathered, thereby improving the relevance and accuracy of threat assessments. Focusing on the African context, a CTI platform tailored for organizations incorporates data from social media platforms like Twitter, enhancing situational awareness despite not specifically targeting disinformation~\cite{africantip}. Furthermore, a platform designed for correlating CTI from Open-Source Intelligence (OSINT) sources demonstrates the effectiveness of aggregating publicly available data to identify potential threats~\cite{correlationosinttip}. 

Leveraging machine learning techniques, the inTIME framework~\cite{machinelearningtip} automates the gathering and analysis of web data for CTI, showcasing the potential of artificial intelligence in enhancing cybersecurity measures. Similarly, ThreatWise AI~\cite{aiholistictiframework} integrates AI and machine learning in a framework to enrich and analyze CTI data by using MISP objects and other external sources with a novel pipeline. Additionally, the TSTEM platform~\cite{inthewildtip} employs cognitive computing to collect CTI from diverse online sources, including social media and websites, facilitating real-time threat detection and analysis. These initiatives underscore the critical role of structured and interoperable CTI platforms in strengthening cybersecurity defenses across various sectors.

On the contrary, CTI approaches related to disinformation incidents are limited. Some public databases and works have emerged to gather disinformation incidents in large quantities. For example, EUvsDisinfo~\cite{euvsdisinfoEUvsDisinfoDetecting}, managed by the East Stratcom Task Force, gathers over 18,200 reports on disinformation incidents with summaries and fixed properties. Similarly, Disinfodex~\cite{disinfodex}, supported by the Harvard Berkman-Klein Center, documents 379 disinformation campaigns on platforms like Google and Facebook, including details about removed resources and policy violations. Other initiatives, such as the Media Manipulation Casebook~\cite{mediamanipulationMediaManipulation} with 36 entries and the DFRLab’s Foreign Interference Attribution Tracker (FIAT)~\cite{interference2024Interference2024} with 86 entries, expand on these efforts by coding disinformation campaigns with relevant variables and visualizing trends. Fulde-Hardy~\cite{fulde} performs an analysis of election-related disinformation campaigns from 2014 to 2024, employing the DISARM framework to model the analyzed incidents, resulting in a rich dataset with 81 campaigns. However, these disinformation-based repositories are not implementing homogeneous and standardized sharing methodologies for CTI, making it difficult to programmatically consume that intelligence. 

Given the success of community-driven threat exchange solutions in cybersecurity, a similar approach could be applied to manage disinformation campaigns. Recent initiatives, such as the Defending Against Deception Common Data Model (DAD-CDM) project~\cite{dadcdmHomeDADCDM} by OASIS in 2023, aim to introduce a common data model for normalizing and sharing information on disinformation campaigns using the STIX standard and leveraging advances from the DISARM framework. Additionally, OpenCTI~\cite{filigranOpenCTIFiligran}, a popular open-source solution by Filigran for threat exchange and CTI management, can serve as a merging point for different CTI feeds. It already features a DISARM connector~\cite{githubConnectorsexternalimportdisarmframeworkMaster} that enables the platform to build reports with DISARM’s TTPs and better represent disinformation threats. Nevertheless, this connector remains basic, primarily aimed at generating reports within OpenCTI rather than enabling the automatic ingestion of disinformation incidents from databases or datasets.

In reviewing the landscape of existing CTI platforms, we identify the need for a dedicated solution that addresses disinformation-specific challenges. While platforms like MISP, OpenCTI, EclecticIQ, and OTX have proven instrumental in managing cybersecurity threats, their design primarily revolves around traditional cybersecurity concepts such as IoCs and threat actor mapping. Although some solutions offer limited extensions for disinformation, these features are often secondary and lack the focused structure required for effective disinformation analysis and exchange. This paper explores this critical gap by adopting a purpose-built and agnostic approach, specifically designed to model, analyze, and share intelligence about disinformation campaigns, also tackling the gaps in current unstructured disinformation repositories. Grounded in the DISARM framework and fully aligned with the STIX2 standard, our framework ensures interoperability with other CTI platforms while maintaining a lightweight and scalable design.

\section{Modeling of disinformation incidents}
\label{sec:disinfomodeling}
For CTI, accurately modeling threats is essential for formal and homogeneous analysis, sharing, and response. Similar to how cyberattacks are deconstructed using cyber kill chains, disinformation attacks require structured modeling to capture their phases and strategies. This enables a common understanding and translation into standardized formats like STIX2, fostering interoperability and automation in combating information threats jointly in both countries and organizations.

\subsection{Comparison of disinformation frameworks}

A recent article~\cite{disinfomodel} reviews the pros and cons of disinformation-based schemes and taxonomies, having different perspectives and applications. Table \ref{tab:discussionFrameworks} presents a summary of the frameworks considered for modeling disinformation incidents. 
This section provides a comparative analysis of five prominent frameworks: DISARM, SCOTCH, BEND, ABCDE, and ALERT. These frameworks vary in their focus, design, and applicability, offering diverse approaches to understanding and mitigating disinformation campaigns

\subsubsection{Framework description}
The Disinformation Analysis and Risk Management (DISARM) framework~\cite{disarm}, proposed by the DISARM Foundation, is a comprehensive model inspired by cybersecurity practices. It employs the MITRE ATT\&CK model and Cyber Kill Chain analogy, which are widely recognized in the cybersecurity domain. 
DISARM outlines a four-stage matrix (Plan, Prepare, Execute, and Assess) with specific TTPs, which offers a systematic and structured approach to modeling disinformation behaviors. Additionally, it provides an STIX2 mapping to codify the TTPs effectively with a standardized language for sharing threat intelligence.

The Source, Channel, Objective, Target, Composition and Hook (SCOTCH) framework~\cite{scotch}, developed by the Atlantic Council, is a high-level framework for understanding disinformation campaigns, particularly focusing on rapidly assessing influence operations by looking to a more abstract layer and analyzing the source, channel, objective, target, composition and hook. It offers insights into disinformation classification, making it a valuable resource for practitioners who need actionable guidance.

The BEND framework~\cite{bend}, created by Carnegie Mellon University in collaboration with the US Army, provides a structured framework for identifying and responding to disinformation threats. It is notable for including quantitative analysis, disinformation classification, and countermeasures, providing a more technical and measurable approach compared to others. However, it does not include interoperable codification capabilities, which may limit its compatibility with standardized intelligence-sharing formats.

The Actor, Behavior, Content,
Degree, and Effect (ABCDE) framework~\cite{Pamment2020}, proposed by the Carnegie Endowment for International Peace, takes a more conceptual approach, concentrating on actor analysis and qualitative assessments. While it provides useful insights into the motivations and behaviors of actors involved in disinformation campaigns, it lacks features such as incident stages and codification capabilities, making it less actionable in practice.

Finally, the Actor, Lever, Effects, and Response Taxonomy (ALERT) framework~\cite{alert}, developed by QUT Business School, the University of Melbourne, and IDSA, offers a broad framework for understanding disinformation campaigns. It presents a taxonomy based on actors, lever, effects and responses, aiming to help security practitioners and policymakers in analyzing disinformation attacks in information systems. However, ALERT is more conceptual than operational, making it better suited for high-level strategic analyses rather than tactical applications.

\subsubsection{Comparative analysis}

\begin{table}[]
\begin{adjustbox}{width=\textwidth,center}
\begin{threeparttable}
    \begin{tabular}{ p{5.25cm}  C{3.25cm}  C{3.25cm}  C{3.25cm}  C{3.25cm}  C{3.25cm} }
\toprule
 \textbf{Features} &
  \textbf{\faStar\ DISARM} &
  \textbf{SCOTCH} &
  \textbf{BEND} &
  \textbf{ABCDE} &
  \textbf{ALERT} \\ \midrule
  Proposed by &
  DISARM Foundation
  &
  Atlantic Council
  &
  Carnegie Mellon and US Army
  &
  Carnegie Endowment for International Peace
  &
  QUS Business School, University of Melbourne and IDSA
  \\ \midrule
Disinformation classification &
  X&
  X&
  X&
  X&
  X
   \\ \midrule
Use case examples &
  X&
  X&
  X&
  X&
  X
   \\ \midrule
Actors analysis &
  - &
  X &
  X &
  X &
  X
  \\ \midrule

Countermeasures &
   X&
   -&
   X&
   X&
   X
   \\ \midrule
Quantitative analysis &
   - & 
   - & 
   X &
   - &
   - \\ \midrule
Supported by &
  EU, OTAN, ONU&
  -&
  -&
  EU&
  -
   \\ \midrule
Codification capabilities &
  STIX2 &
  -&
  TSV&
  - &
  -
   \\ \midrule
   Stages &
   Plan, Prepare, Execute, Assess&
   -&
   Framework workflow&
   -&
   -
   \\ \midrule
Cyber analogy & 
   MITRE ATT\&CK and Cyber Kill Chain&
    -&
    -&
    -&
    -\\ 
   \bottomrule
\end{tabular}%
\end{threeparttable}

\end{adjustbox}
\caption{Summary of frameworks analysed (adapted from our recent publication~\cite{disinfomodel})}
\label{tab:discussionFrameworks}

\end{table}

The ability to characterize and model disinformation incidents is the base property of all the frameworks. This capability is particularly useful for organizations aiming to analyze the diversity of disinformation campaigns. However, all of them have their particularities.

The inclusion of real-world examples helps bridge the gap between theory and application. All the analyzed frameworks—DISARM, SCOTCH, BEND, ABCDE, and ALERT—provide use case examples, making them valuable for practitioners seeking to understand their practical implementation. However, DISARM and BEND excel in demonstrating how their methodologies can be applied to real-world scenarios, offering detailed illustrations that enhance their utility.

Understanding the roles and motivations of actors is a key strength of several frameworks. ABCDE, SCOTCH, BEND, and ALERT emphasize actor analysis, providing tools for identifying and examining the key players involved in disinformation campaigns. However, DISARM does not explicitly offer actor-focused analysis, as it is more centered on technical and procedural aspects.

Developing effective countermeasures is a critical aspect of disinformation frameworks. DISARM, BEND, ABCDE and ALERT stand out in this regard by explicitly including countermeasure planning within their models. DISARM, in particular, integrates a mapping between used techniques and the countermeasures to tackle it, providing a direct and actionable approach. ABCDE and ALERT also include countermeasure considerations but they are limited to recommendations for very open scenarios, contrary to the directness offered by DISARM. Conversely, SCOTCH does not explicitly include countermeasures, limiting their operational relevance.

Quantitative analysis is a valuable feature for organizations seeking measurable insights into disinformation campaigns. BEND incorporates quantitative methodologies, enabling users to evaluate the impact and scale of campaigns. However, contrary to some perceptions, DISARM does not explicitly integrate quantitative analysis into its framework, focusing instead on TTPs and technical interoperability. This feature is also absent in SCOTCH, ABCDE, and ALERT, which rely more heavily on high-level assessments.

Codification capabilities enhance interoperability with existing systems and standards. DISARM is the only framework to adopt STIX2, a widely used standard for threat intelligence sharing, ensuring seamless integration into cybersecurity workflows. BEND supports TSV formatting for use with ORA-PRO software, providing some degree of codification but lacking the standardization advantages of STIX2. SCOTCH, ABCDE, and ALERT do not offer codification features, limiting their ability to integrate into technical ecosystems.

The methodologies and processes defined by the frameworks vary significantly in their structure and detail. DISARM outlines a comprehensive four-stage methodology—Plan, Prepare, Execute, and Assess— with TTPs rooted in cybersecurity practices. BEND adopts a workflow-based approach that focuses on maneuvering narratives and social networks, while SCOTCH, ABCDE, and ALERT remain high-level conceptual frameworks, offering general guidance rather than specific methodologies.

Cybersecurity analogies, such as MITRE ATT\&CK and the Cyber Kill Chain, provide valuable context for addressing disinformation in technical settings. Among the analyzed frameworks, only DISARM incorporates these analogies, making it uniquely suited for organizations familiar with cybersecurity practices. The other frameworks do not draw on these analogies, adopting broader approaches that may lack the precision needed for technical integration.

After this comparison, DISARM emerges as the most comprehensive model, combining the use of TTPs with codification capabilities and a structured methodology. 
SCOTCH and ALERT, while less technical, provide valuable tools for actor analysis and classification, making them useful for strategic and conceptual analyses. BEND stands out for its quantitative focus, offering measurable tools for analyzing disinformation threats and their impact. ABCDE, on the other hand, offers a high-level conceptual framework that is valuable for qualitative assessments but lacks actionable features for operational use in our context.

\subsection{Selection of DISARM as a reference framework} \label{sec:disarmselection}
The DISARM framework~\cite{disarm} integrates the concept of TTPs to model the behaviors and actions in disinformation attacks. It merges tools like the MITRE ATT\&CK matrix or the Cyber Kill Chain and adapts them to enable a rich description of disinformation incidents by proposing a large set of DISARM in a matrix, detailed in Section~\ref{sec:matrix}. Additionally, the project provides an initial approach~\cite{githubGitHubDISARMFoundationDISARMSTIX2} to model attack techniques in STIX2, offering a direct mapping of disinformation attack techniques to the \texttt{AttackPattern} STIX object type. It also includes an official OpenCTI connector for integrating its TTPs matrix into the platform, enabling visualization and correlation of incidents. The aforementioned capabilities and applications demonstrate that the DISARM framework provides a clear cybersecurity perspective, making it an ideal choice for modeling disinformation incidents within the threat intelligence platform developed in this work.

Moreover, the utility of DISARM has been endorsed by several official EU bodies, including FIMI-ISAC~\cite{fimiisac-findings}, the EEAS~\cite{eeas1, eeas2}, ENISA~\cite{enisathreat} or Hybrid CoE~\cite{HybridCoE2022}. It is also employed in disinformation-related reports from ADAC.IO~\cite{adacdisarm}, the ATHENEA project~\cite{athenea}, the EDMO~\cite{edmoelections} or EU Disinfolab~\cite{disinfolabeufimi}, further demonstrating the increasing adoption of this framework.

\subsection{DISARM TTP Matrix}
\label{sec:matrix}
The core of the DISARM framework is its MITRE ATT\&CK-like matrix, which can be visualized online\footnote{\url{https://disarmframework.herokuapp.com}}. The matrix permits the decomposition of any incident in phases with associated tactics and techniques. In the following, we formally define the main concepts of the matrix and apply them to a real-world influence operation within the Russia-Ukranian war for clear comprehension. As this example will also showcase the rest of the paper, we provide some context.

\begin{tcolorbox}[colframe=black, colback=gray!10, boxrule=0.5pt]
The \textit{Ukraine Re-sold French Howitzers} (URFH) disinformation incident involved claims that Ukraine had sold CAESAR howitzers—supplied by France as military aid—on the black market. These allegations were propagated by Russian-affiliated media and Telegram channels in July 2022, supported by fabricated evidence and unverifiable reports. The narrative aimed to undermine trust in Western military support for Ukraine and to portray the aid as being misused. Despite lacking credible evidence, the disinformation gained traction within pro-Russian circles, showcasing the manipulation of information to influence public perception during the Russia-Ukraine war~\cite{mediumRussiaPromoted}.
\end{tcolorbox}

In this sense, to the eyes of the DISARM framework, the operation can be matched to the matrix and its elements which are described next. Table \ref{tab:disarmred} illustrates the application of this matrix to the defined use case, supporting the description of the DISARM elements:

\begin{table}[h!]
\begin{threeparttable}
\centering
\begin{tabularx}{\textwidth}{ p{2cm} X  X }
\toprule
\multicolumn{3}{c}{\textbf{Phase: \texttt{PLAN}}} \\ 
\midrule
 {\textbf{Tactic}} & {\textbf{Technique}} & {\textbf{Rationale}} \\ 
\midrule
\multirow{1}{*}{\makecell[l]{\parbox{\linewidth}{\texttt{TA02: Plan Objectives}}}}
   & \texttt{T0002: Facilitate State Propaganda}  & \textit{Coordinating volunteers to disseminate messages benefiting Russia.}  \\ 
\midrule
\multicolumn{3}{c}{\textbf{Phase: \texttt{PREPARE}}} \\ 
\midrule
 {\textbf{Tactic}} & {\textbf{Technique}} & {\textbf{Rationale}} \\ 
\midrule
\multirow{2}{*}{\makecell[l]{\parbox{\linewidth}{\texttt{TA06: Develop Content}}}}
   & \texttt{T0019.001: Create fake research}  & \textit{``Experts"\ claiming that Russia replicated the howitzers} \\ \cmidrule{2-3}
   & \texttt{T0040: Demand insurmountable proof} & \textit{Russian media reframing French' official versions} \\
\midrule
\multirow{4}{*}{\makecell[l]{\parbox{\linewidth}{\texttt{TA07: Channels \& Affordances}}\vspace{1cm}}}
   & \texttt{T0043: Chat apps} & \textit{Telegram use} \\ \cmidrule{2-3}
   & \texttt{T0104: Social Networks} & \textit{Twitter use} \\ \cmidrule{2-3}
   & \texttt{T0111: Traditional Media} & \textit{News in pro-Russian outlets} \\
\midrule
\multicolumn{3}{c}{\textbf{Phase: \texttt{EXECUTE}}} \\
\midrule
 {\textbf{Tactic}} & {\textbf{Technique}} & {\textbf{Rationale}} \\ 
\midrule
\texttt{TA08: Conduct Pump Priming} & \texttt{T0045: Use fake experts}  & \textit{``Experts"\ claiming that Russia replicated the howitzers} \\
\midrule
\multirow{3}{*}{\makecell[l]{\parbox{\linewidth}{\texttt{TA09: Deliver content}}}} 
   & \texttt{T0115.003: One-Way Direct Posting} & \textit{Telegram channels to disseminate} \\ \cmidrule{2-3}
   & \texttt{T0119: Cross-Posting} & \textit{Using news sites, Telegram, Twitter and other platforms} \\ \cmidrule{2-3}
   & \texttt{T0117: Attract Traditional Media} & \textit{News reaching mainstream media} \\
\bottomrule

\end{tabularx}
\caption{DISARM phases, tactics and techniques detected in the ``Ukraine Re-sold French Howitzers" disinformation campaign by Russian actors in the Russia-Ukraine war.}
\label{tab:disarmred}
\end{threeparttable}
\end{table}

\begin{enumerate}
    \item \textbf{Phase}: The most abstract grouping, representing sequential stages of an influence campaign by combining related tactics. There are four phases, including 1) \texttt{PLAN} (defining objectives and strategies), 2) \texttt{PREPARE} (creating and organizing assets), 3) \texttt{EXECUTE} (deploying and amplifying content), and 4) \texttt{ASSESS} (evaluating performance and persistence).

     \begin{tcolorbox}[colframe=black, colback=gray!10, boxrule=0.5pt]
     In the URFH incident, the first three phases of \texttt{PLAN}, \texttt{PREPARE} and \texttt{EXECUTE} can be inferred, but the last one of \texttt{ASSESS} is not intuitively interpretable by the analyst.
    \end{tcolorbox}
    
    \item \textbf{Tactic}: Specific strategy that can be deployed in a particular Phase to achieve the campaign effects. The \texttt{PLAN} phase includes three possible tactics: \texttt{Plan Strategy}, \texttt{Plan Objectives}, and \texttt{Target Audience Analysis}, which outline the strategic groundwork. The \texttt{PREPARE} phase encompasses six tactics: \texttt{Develop Narratives}, \texttt{Develop Content}, \texttt{Establish Social Assets}, \texttt{Establish Legitimacy}, \texttt{Microtarget} and \texttt{Select Channels and Affordances}, focusing on operational readiness. The \texttt{EXECUTE} phase groups six tactics such as \texttt{Conduct Pump Priming}, \texttt{Deliver Content}, \texttt{Maximize Exposure}, \texttt{Drive Online Harms},  \texttt{Drive Offline Activity} and \texttt{Persist in the Information Environment}, ensuring active dissemination and impact. Lastly, the \texttt{ASSESS} phase includes only the tactic of \texttt{Assess Effectiveness}, emphasizing evaluation and refinement of the campaign's outcomes. As shown in Table \ref{tab:disarmred}, DISARM universally tags each tactic with a numerical unambiguous identifier.
    
    \begin{tcolorbox}[colframe=black, colback=gray!10, boxrule=0.5pt]
    In the URFH use case, Russia would \texttt{Plan Objectives} during the \texttt{PLAN} phase, \texttt{Develop Content} and \texttt{Select Channels \& Affordances} during the \texttt{PREPARE} phase, and \texttt{Conduct Pump Priming} and \texttt{Deliver Content} during the \texttt{EXECUTE} phase.
    \end{tcolorbox}
    
    \item \textbf{Technique}: Specific fine-grained action deployed in the real world to complete a tactic. A tactic can have multiple techniques, one may be associated with multiple tactics, and some have sub-techniques for further detail. The DISARM framework covers a wide range of dozens of techniques to interpret any movement of any investigated operation, as mentioned next.
    
\begin{tcolorbox}[colframe=black, colback=gray!10, boxrule=0.5pt]
    In the URFH campaign, the actor begins in the \texttt{PLAN} phase with the \texttt{Plan Objectives} tactic, utilizing \texttt{Facilitate State Propaganda} to organize volunteers and disseminate messages favorable to their agenda. Moving to the \texttt{PREPARE} phase, the \texttt{Develop Content} tactic is employed through \texttt{Create Fake Research} and \texttt{Demand Insurmountable Proof}, aimed at discrediting opposing narratives and creating doubt about official information. Concurrently, the \texttt{Select Channels \& Affordances} tactic leverages \texttt{Chat Apps}, \texttt{Social Networks}, and \texttt{Traditional Media} to ensure targeted and broad distribution of the fabricated content. Finally, in the \texttt{EXECUTE} phase, the actor applies the \texttt{Conduct Pump Priming} tactic using \texttt{Use Fake Experts} to lend false credibility to their claims. They further amplify the message through the \texttt{Deliver Content} tactic, employing \texttt{Cross-Posting}, \texttt{One-Way Direct Posting}, and \texttt{Attract Traditional Media} to maximize reach across various platforms and audiences.
\end{tcolorbox}

\end{enumerate}


As a conclusion, the DISARM framework provides a method to characterize and understand a complex influence operation universally.




\section{STIX2 codification of DISARM-modeled disinformation incidents} \label{sec:mapping}

For the solution to be CTI-compatible, the real-world disinformation incident modeled with DISARM shall be translated to STIX2 objects, providing computational interoperability between connectors using this widely adopted threat intelligence data format.  

STIX2~\cite{stix2} structures threat intelligence information in a standardized JSON-based format, traditionally focusing on cyberattacks. It organizes data into a bundle of interconnected objects, each representing predefined aspects of an incident, such as observed behaviors, threat actors, tools, or techniques. This structured approach ensures a consistent representation and enables seamless integration between systems.

However, although there are standardized ways of transforming cybersecurity knowledge to STIX2 objects, there are no guidelines for representing disinformation incidents yet. Therefore, we have conceptualized a way to abstract the nature of disinformation incidents to fit them in the already available STIX2 objects, providing an equivalency between a disinformation incident and a cybersecurity incident. This is also powerful, as it supports the integration and correlation in the same domain and common language of information and cyber threats, which is important for today's context.

\subsection{Disinformation entities through STIX Domain Objects (SDOs)}
Firstly, the STIX Domain Objects (SDOs) define specific concepts usually found in the CTI ecosystem~\cite{stix2}. As shown in Table \ref{tab:sdos}, we map the details related to a disinformation incident to particular standardized STIX objects as follows:

\begin{table}[htbp]
    \centering
    \begin{threeparttable}

    \begin{tabularx}{\textwidth}{ p{2cm}  l  X }
        \toprule
        \multicolumn{3}{c}{\textbf{STIX Domain Objects (SDOs)}} \\ 
        \midrule
        \textbf{Property}     & \textbf{STIX2 object}  & \textbf{Rationale}     \\
        \midrule
        \midrule
        \textit{Incident}         & \texttt{IntrusionSet}  & Group of actions done by some entity          \\ \midrule
        \textit{Actor}         & \texttt{ThreatActor}     &  Author of the incident        \\ \midrule
        \textit{Technique}  & \texttt{AttackPattern} & DISARM technique launched \\ \midrule
        \textit{Country} & \texttt{Location} & Geographic point of the targeted region  \\ 
        \midrule
    \end{tabularx}
    \caption{Mapping between disinformation properties (nodes) and STIX2 object types}
    \label{tab:sdos}
        \end{threeparttable}
\end{table}

\begin{itemize}
    

\item \textit{Incident}: The core element of a disinformation incident, characterized by key properties such as \texttt{name}, \texttt{description}, and \texttt{first\_seen}. 
It is mapped to an \texttt{IntrusionSet} STIX object, traditionally used to represent a group of cybersecurity activities and resources with shared objectives. 
This aligns well with the strategic and coordinated nature of disinformation incidents. The \texttt{IntrusionSet} serves as the central object characterizing the incident, linking all related entities.

Listing \ref{lst:stixincident} provides a simplified STIX2 representation of the URFH \textit{Incident}. 
The fields \texttt{id}, \texttt{type}, \texttt{created}, \texttt{modified} and  \texttt{spec_version} represent the STIX metadata that define and identify the object itself. 
The remaining fields, such as \texttt{name}, \texttt{description}, \texttt{labels}, and \texttt{first\_seen}, form the payload of the object, containing the core details about the disinformation incident. 

\begin{ccode}[list text=,title={Simplified \texttt{IntrusionSet} SDO representation of the URFH \textit{Incident}}, label=lst:stixincident]
{
  "id": "intrusion-set--76271730-...",
  "type": "intrusion-set",
  "created": "2024-12-25T23:35:11.86288Z",
  "modified": "2024-12-25T23:35:11.86288Z",
  "spec_version": "2.1",
  "name": "Ukraine re-sold French howitzers for profit",
  "description": "Claims that Ukraine had sold CAESAR howitzers...",
  "labels": [ "incident", "disinformation"],
}
\end{ccode}

    \item \textit{Actor}: The entity, whether an organization, group, or individual, is believed to be responsible for orchestrating the \textit{Incident}. It is mapped to a \texttt{ThreatActor} STIX object, which is actually designed to represent the malicious cyberattacker.
    
    Listing \ref{lst:stixactor} contains the STIX2 representation of the \textit{Actor} responsible for the URFH incident. 
    In this representation, the key field is the \texttt{name}, which stores the name of the actor attributed in the source report: Russia. 
    Additionally, the \texttt{threat\_actor\_types} field categorizes the actor as a \texttt{nation-state}, 
    indicating its classification within the threat intelligence ecosystem.
    
    \begin{ccode}[list text=,title={Simplified \texttt{ThreatActor} SDO related to the URFH \textit{Incident}}, label=lst:stixactor]
{
  "id": "threat-actor--7ebead2d-...",
  "type": "threat-actor",
  "created": "2024-12-25T23:27:53.696031Z",
  "modified": "2024-12-25T23:27:53.696031Z",
  "spec_version": "2.1",
  "name": "Russia",
  "labels": ["threat-actor"],
  "threat_actor_types": ["nation-state"]
}
    \end{ccode}

    \item \textit{Technique}: The specific DISARM technique used in the disinformation incident that supported the \textit{Actor} actions to achieve its goals. As Section~\ref{sec:disarmselection} mentions, the DISARM foundation already translated this information to the \texttt{AttackPattern} STIX object for encapsulating the malicious techniques.
    
    Listing \ref{lst:stixattackpattern} presents the STIX2-formatted representation associated with the \textit{Facilitate State Propaganda} DISARM technique employed in the URFH \textit{Incident}. Notice how the \texttt{name} and \texttt{description} fields correspond to the official name and description of the technique \footnote{{https://github.com/DISARMFoundation/DISARMframeworks/blob/main/generated_pages/techniques/T0002.md}}, respectively.  
    The \texttt{kill_chain_phases} field specifies the overarching tactic in the DISARM matrix: \texttt{plan-objectives}, which is utilized by OpenCTI to display the techniques with color-coded visualizations.
    
    In this case, note that the \texttt{created} and \texttt{modified} timestamps differ more than a year from those of the other listed SDOs. This discrepancy arises because these objects were originally created by DISARM in its repository some time ago and the codification process in the platform uses these original
    SDOs instead of creating new ones.

\begin{ccode}[list text=,title={Simplified \texttt{AttackPattern} SDO related with the URFH \textit{Incident}}, label=lst:stixattackpattern]
{
  "id": "attack-pattern--70717452-...",
  "type": "attack-pattern",
  "created": "2023-09-14T20:38:04.999444Z",
  "modified": "2023-09-14T20:38:04.999444Z",
  "created_by_ref": "identity--f1a0f560-...",
  "name": "Facilitate State Propaganda",
  "description": "Organise citizens around pro-state messaging...",
  "external_references": [
    {
      "external_id": "T0002",
      "source_name": "mitre-attack",
      "url": "https://github.com/DISARMFoundation/DISARMframeworks/blob/main/generated_pages/techniques/T0002.md"
    }
  ],
  "kill_chain_phases": [
    {
      "kill_chain_name": "mitre-attack",
      "phase_name": "plan-objectives"
    }
  ],
  ...
}
\end{ccode}
    
    \item \textit{Country}: The world location to which the disinformation attack was targeted to. They are mapped to \texttt{Location} STIX objects as they represent a geographic point.
    
Listing \ref{lst:stixlocation} presents the STIX2-formatted representation of one of the targeted countries identified in the URFH incident: France.  
In this \texttt{Location} SDO, two fields are significant: the \texttt{country} field, which stores the value \texttt{France}, and the \texttt{name} field, which redundantly stores the same value for clarity and identification.

\begin{ccode}[list text=,title={Simplified \texttt{Location} SDO related to the URFH \textit{Incident}}, label=lst:stixlocation]
{
  "id": "location--be5032fd-...",
  "created": "2024-12-25T23:27:52.703244Z",
  "modified": "2024-12-25T23:27:52.703244Z",
  "spec_version": "2.1",
  "country": "France",
  "name": "France",
  "type": "location"
}
\end{ccode}

\end{itemize}

In this context, a disinformation incident can be described using the aforementioned objects. It is important to note that STIX entities are independent of their relationships. This separation is leveraged to flexibly connect entities and expand knowledge, enabling adaptable and extensible modeling through multiple incidents.

\subsection{Disinformation entity relations through STIX Relationship Objects (SROs)}

The STIX Relationship Objects (SROs) link the SDOs and describe the generated CTI \cite{stix2}. As shown in Table \ref{tab:sros}, we define three types of standard STIX relationships that relate two pieces of information (SDOs) through their unique identification (\texttt{id}):

\begin{table}[htbp]
    \centering
    \begin{threeparttable}

    \begin{tabularx}{\textwidth}{ p{4cm}  l  X }
        \midrule
        \multicolumn{3}{c}{\textbf{STIX Relationship Objects (SROs)}} \\
        \midrule
        \textbf{Relationship}     & \textbf{STIX2 object}  & \textbf{Rationale}     \\
        \midrule
        \midrule
        \textit{Incident ${\rightarrow}$ Technique} & \texttt{uses} &  A \textit{Technique} is used in an \textit{Incident} \\ \midrule
        \textit{Incident ${\rightarrow}$ Actor} & \texttt{attributed-to} &  An \textit{Incident} is attributed to some \textit{Actor} \\ \midrule
        \textit{Incident ${\rightarrow}$ Country} & \texttt{targets}  &  An \textit{Incident} targeted to some \textit{Country} \\ 
        \bottomrule
    \end{tabularx}
    \caption{Mapping between disinformation relationships (edges) and STIX2 object types}
    \label{tab:sros}
        \end{threeparttable}
\end{table}

\begin{itemize}
\item \textit{Incident $\overset{uses}{\rightarrow}$ Technique}: Represents the relationship between a \textit{DISARM Technique} and the \textit{Incident} in which it was employed. Typically, an \textit{Incident} involves multiple \textit{Techniques}, resulting in many such relationships.  

Listing \ref{lst:stixsrotech} shows the STIX2 representation of the URFH disinformation technique. The \texttt{relationship\_type} field is set to \texttt{uses}, aligning with our definition. The \texttt{source\_ref} field references the \texttt{id} of the \texttt{IntrusionSet} representing the URFH \textit{Incident}, while the \texttt{target\_ref} field points to the \texttt{id} of the \texttt{AttackPattern} representing the \textit{Facilitate State Propaganda} DISARM \textit{Technique}.

\begin{ccode}[list text=,title={Simplified \texttt{uses} SRO related to the URFH \textit{Incident}}, label=lst:stixsrotech]
{
  "id": "relationship--1dce08d4-3650-4f78-8d55-1a08055ffbf3",
  "relationship_type": "uses",
  "source_ref":"intrusion-set--76271730-6e05-51f0-bf4c-6a7c7b53d9b0",
  "target_ref":"attack-pattern--70717452-f7e3-4ce8-956f-39a4d34c5cfb",
  "type": "relationship",
  ...
}
\end{ccode}

\item \textit{Incident $\overset{attributed\ to}{\rightarrow}$ Actor}: Represents the relationship between an \textit{Actor} and the \textit{Incident} attributed to it.  

Listing \ref{lst:stixsroactor} shows the STIX2 representation of the URFH attribution. The \texttt{relationship\_type} field is set to \texttt{attributed-to}. The \texttt{source\_ref} field references the \texttt{id} of the \texttt{IntrusionSet} representing the URFH \textit{Incident}, and the \texttt{target\_ref} field points to the \texttt{id} of the \texttt{ThreatActor} representing the Russia \textit{Actor}.

\begin{ccode}[list text=,title={Simplified \texttt{attributed-to} SRO related to the URFH \textit{Incident}}, label=lst:stixsroactor]
{
  "id": "relationship--dd7da138-6850-4b6b-ae0f-8f20c2502882",
  "relationship_type": "attributed-to",
  "source_ref":"intrusion-set--76271730-6e05-51f0-bf4c-6a7c7b53d9b0",
  "target_ref":"threat-actor--7ebead2d-9a79-505f-8998-026100724eab",
  "type": "relationship",
  ...
}
\end{ccode}

\item \textit{Incident $\overset{targets}{\rightarrow}$ Country}: Represents the relationship between a \textit{Country} and the \textit{Incident} that targeted it.  

Listing \ref{lst:stixsrolocation} shows the STIX2 representation of the URFH target. The \texttt{relationship\_type} field is set to \texttt{targets}, indicating the targeting relationship. The \texttt{source\_ref} field refers to the \texttt{id} of the \texttt{IntrusionSet} representing the URFH \textit{Incident}, and the \texttt{target\_ref} field points to the \texttt{id} of the \texttt{Location} object representing the France \textit{Country} targeted in URFH incident.

\begin{ccode}[list text=,title={Simplified \texttt{targets} SRO related to the URFH \textit{Incident}}, label=lst:stixsrolocation]
{
  "id": "relationship--c476d1ee-1c33-4989-a51c-3dd4ef64dcf5",
  "relationship_type": "targets",
  "source_ref":"intrusion-set--76271730-6e05-51f0-bf4c-6a7c7b53d9b0",
  "target_ref":"location--be5032fd-0b5c-5170-beb7-c7b499afa4bd",
  "type": "relationship"
  ...
}
\end{ccode}

\end{itemize}

\begin{figure}
      \centering
      \includegraphics[width=.98\textheight, angle=-90]{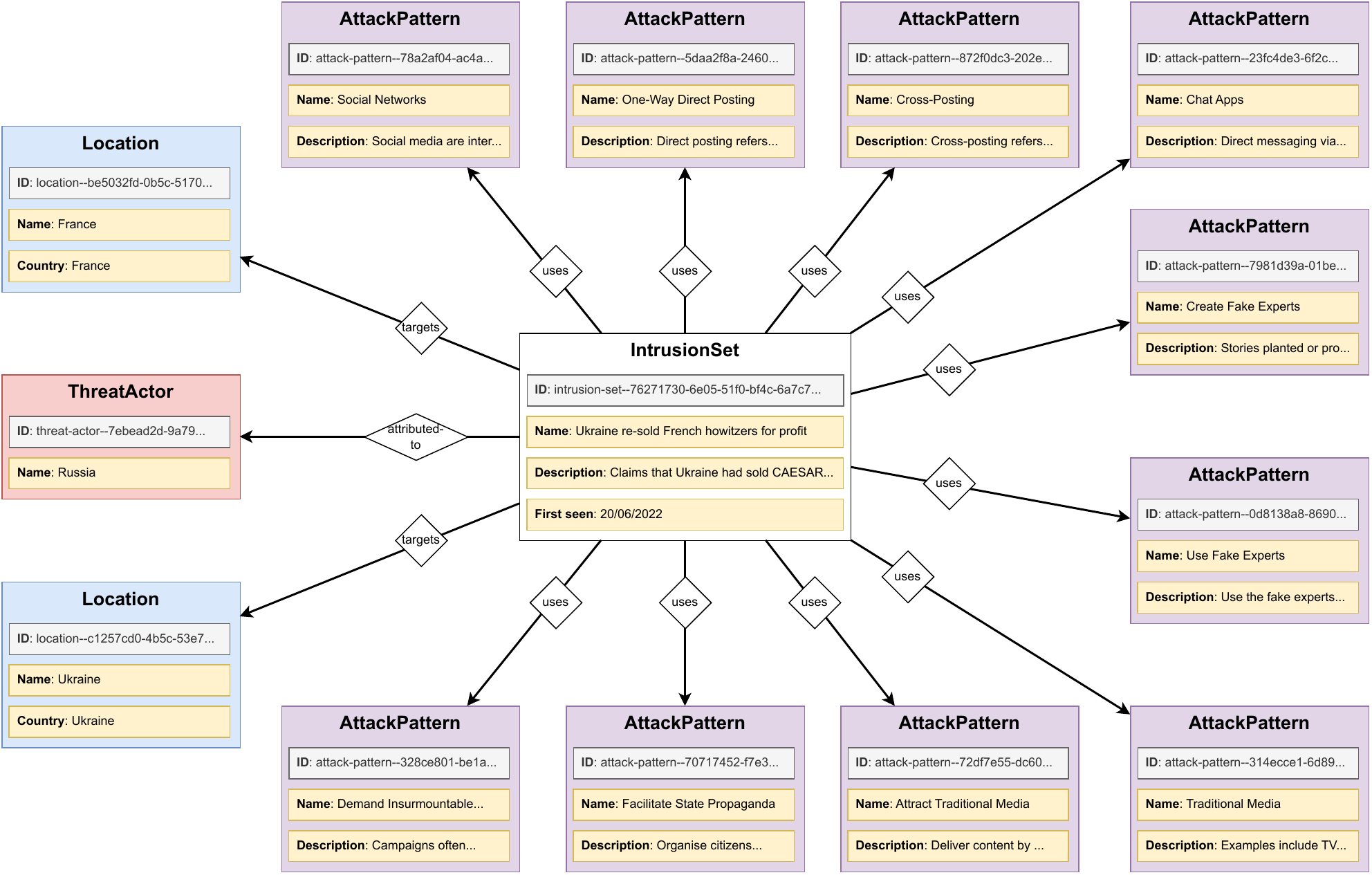}
      \caption{Graph representation of the STIX Bundle representing the modeled URFH disinformation incident}
      \label{fig:stixrepresentation}
  \end{figure}

To sum up, these  STIX2 SDOs and SROs objects constitute standard representations of DISARM-modeled incidents. In order to be exchanged between CTI peers, they are encapsulated in a STIX2 Bundle, a container used to package and share multiple STIX objects~\cite{stix2}. Visually, the STIX2 Bundle can be seen as a graph in Figure \ref{fig:stixrepresentation}. The corresponding simplified, machine-readable STIX2 Bundle object is shown in Listing~\ref{lst:stix2example}, and the full version is available in the project repository\footnote{{https://github.com/CyberDataLab/disinfox/blob/main/backend/data/urfh_incident.json}}.

\begin{ccode}[list text=,title={Simplified STIX2 bundle of uploaded disinformation incident}, label=lst:stix2example]
{
  "id": "bundle--3351770d-0656-4b3b-862f-6e81742669a3",
  "type": "bundle"
  "objects": [
  {
      "description": "Claims that Ukraine had sold CAESAR howitzers...",
      "first_seen": "2022-06-20T00:00:00Z",
      "id": "intrusion-set--76271730-6e05-51f0-bf4c-6a7c7b53d9b0",
      "name": "Ukraine re-sold French howitzers for profit",
      "type": "intrusion-set",
      ...
    },
    {
      "id": "threat-actor--7ebead2d-9a79-505f-8998-026100724eab",
      "name": "Russia",
      "type": "threat-actor",
      ...
    },
    {
      "country": "France",
      "id": "location--be5032fd-0b5c-5170-beb7-c7b499afa4bd",
      "name": "France",
      "type": "location"
      ...
    },
    {
      "created_by_ref": "identity--f1a0f560-2d9e-4c5d-bf47-7e96e805de82",
      "description": "Organise citizens around pro-state messaging. Coordinate paid or volunteer groups to push state propaganda.",
      "external_references": [
        {
          "external_id": "T0002",
          "source_name": "mitre-attack",
          "url": "https://github.com/DISARMFoundation/DISARMframeworks/blob/main/generated_pages/techniques/T0002.md"
        }
      ],
      "id": "attack-pattern--70717452-f7e3-4ce8-956f-39a4d34c5cfb",
      "name": "Facilitate State Propaganda",
      "type": "attack-pattern",
    },
    {
      "id": "relationship--1dce08d4-3650-4f78-8d55-1a08055ffbf3",
      "relationship_type": "uses",
      "source_ref": "intrusion-set--76271730-6e05-51f0-bf4c-6a7c7b53d9b0",
      "target_ref": "attack-pattern--70717452-f7e3-4ce8-956f-39a4d34c5cfb",
      "type": "relationship",
      ...
    },
    {
      "id": "relationship--c476d1ee-1c33-4989-a51c-3dd4ef64dcf5",
      "relationship_type": "targets",
      "source_ref": "intrusion-set--76271730-6e05-51f0-bf4c-6a7c7b53d9b0",
      "target_ref": "location--be5032fd-0b5c-5170-beb7-c7b499afa4bd",
      "type": "relationship"
      ...
    }
    ...
  ],
}
\end{ccode}

\section{DISINFOX: DISINFOrmation eXchange Threat Intelligence architecture} \label{sec:design}  

Having described the modeling and representation of disinformation threat intelligence, the \DISINFOX architecture provides comprehensive, end-to-end support for sharing disinformation incidents. It encompasses the entire process, from uploading incidents in computational language to a centralized server, to the consumption of intelligence by client-side applications.

\subsection{Design of the DISINFOX architecture}

\begin{figure}[h!]
    \centering
    \includegraphics[width=0.5\linewidth]{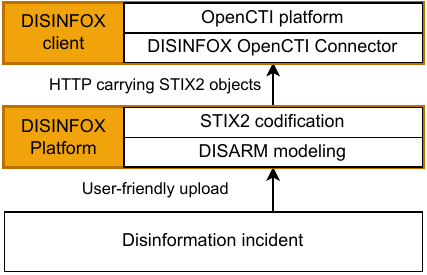}
    \caption{Technological stack of the DISINFOX architecture}
    \label{fig:techstack}
\end{figure}

The \DISINFOX architecture is inspired by well-established deployment models of traditional cybersecurity OTX schemes~\cite{TOUNSI2018212}. It is designed to handle real-world disinformation incidents originating from diverse sources, such as individual initiatives, news sites, or government reports. Figure~\ref{fig:techstack} illustrates the technological stack, showcasing the process from uploading incidents to the platform to their integration within a CTI system. 

The \DISINFOX technological stack features two main components:

\begin{itemize}
    \item \textbf{\DISINFOX platform}: The \DISINFOX platform serves as the centralized repository for standardized, disinformation-based knowledge, providing a persistent source of intelligence and user-friendly management. It ingests disinformation incidents using a two-phase pipeline:
    \begin{enumerate}
        \item \textit{DISARM Modeling}: Applied to represent the techniques used in each incident, as described in Section~\ref{sec:matrix}. Out of the DISARM framework, complementary details such as actor names, affected countries and other contextual data are also ingested.
        
        \item \textit{STIX2.1 Representation}: The extended model of the disinformation incident is transformed into STIX2.1 format, generating SDOs and SROs and inserting them into the database, as detailed in Section~\ref{sec:mapping}. This transformation ensures a standardized and machine-readable representation of the incident.
    \end{enumerate}
    
    \item \textbf{\DISINFOX clients}: They are responsible for consuming and operationalizing disinformation-related knowledge. This paper introduces a custom \DISINFOX OpenCTI Connector integrated with the OpenCTI platform. The \DISINFOX OpenCTI Connector retrieves STIX2-encoded incidents from the \DISINFOX platform and imports them into OpenCTI, enabling visualization and correlation with other CTI objects. 
    Nevertheless, the \DISINFOX client could be other CTI consumers by implementing the corresponding HTTP API based on STIX2.
\end{itemize}

In the following section, we describe the implementation of the \DISINFOX architecture.


\subsection{Implementation of the DISINFOX architecture}

To ensure scalability, flexibility, and integration with existing CTI platforms, \DISINFOX follows a service-oriented architecture, as illustrated in Figure~\ref{fig:architecture}. The architecture comprises multiple interacting components, each responsible for a distinct function within the intelligence lifecycle:

\begin{figure}[h]
    \centering
    \includegraphics[width=1\linewidth]{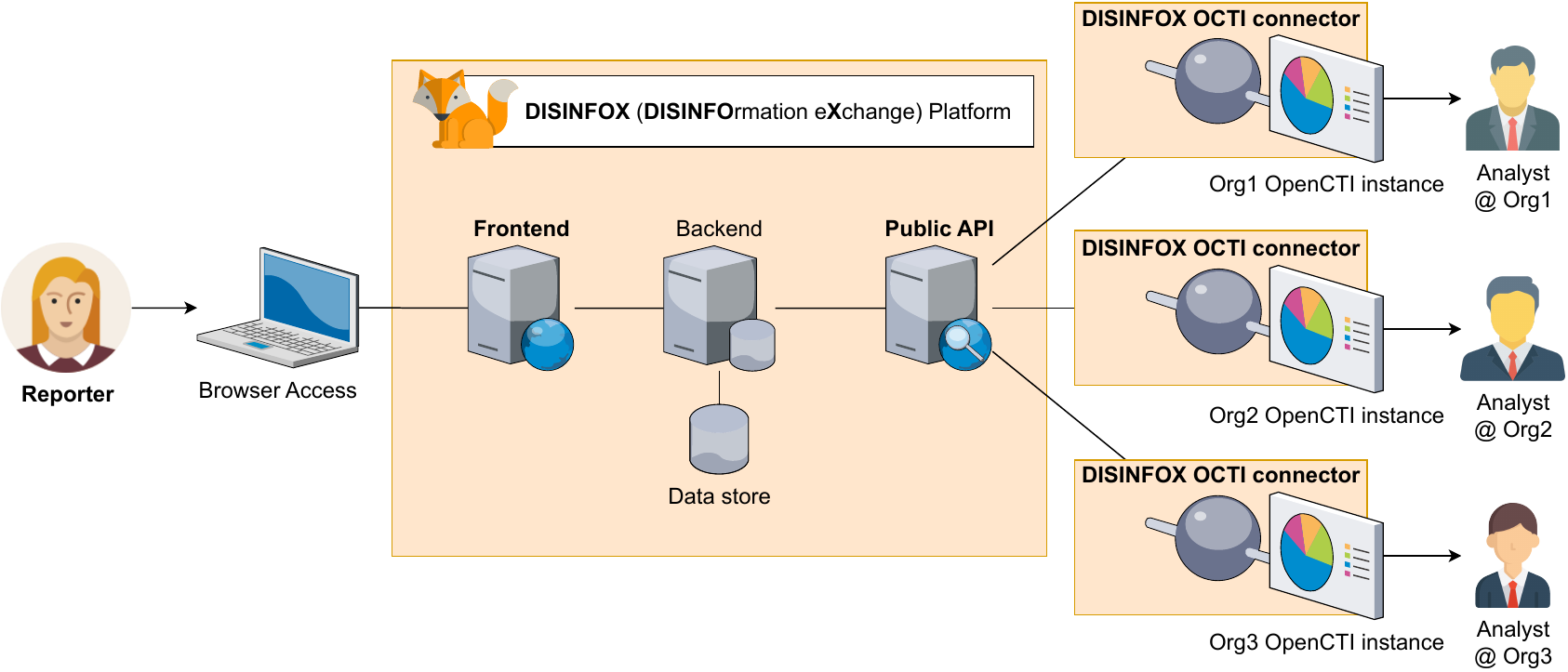}
    \caption{Deployment of the DISINFOX architecture}
    \label{fig:architecture}
\end{figure}

\begin{itemize}
    \item The \textbf{frontend} provides an intuitive web-based interface for non-technical users, facilitating incident submission, visualization, and management. It supports both manual uploads and bulk ingestion of disinformation datasets, ensuring accessibility for a wide range of users.
    \item The \textbf{backend} processes intelligence submissions, validating and structuring incident data according to the STIX2 format. It ensures that each disinformation incident is contextualized, standardized, and interoperable, using DISARM TTPs and the proposed data model. The processed incidents are stored in a document-oriented database, optimized for querying and retrieval by both human analysts and automated systems.
    \item A \textbf{public API} provides structured access to stored incidents, enabling automated retrieval of intelligence updates. External CTI platforms can leverage this API to extract new incidents in real-time, ensuring that disinformation intelligence remains current and actionable.
    
    \item A dedicated \textbf{OpenCTI connector for \DISINFOX} integrates \DISINFOX data directly into OpenCTI to validate interoperability. This connector retrieves structured incidents thanks to the public API and merges them with existing cybersecurity intelligence in OpenCTI, enabling joint analysis of cyber and disinformation incidents.
\end{itemize}

\subsection{Incident lifecycle overview and validation}

The \DISINFOX architecture enables the exchange of threat intelligence. Particularly, disinformation incidents follow a structured pipeline from ingestion to intelligence dissemination. 

Figure~\ref{fig:lifecycle} outlines this process, demonstrating how incidents transition from initial detection to structured intelligence available in CTI platforms. The process begins when a reporter identifies a disinformation campaign (Step 1) and submits key DISARM details (Step 2). The platform validates and structures the submission, transforming it into STIX2 objects following a predefined mapping (Step 3).  Once stored in the centralized database, incidents are retrievable through multiple channels (Step 4):

\begin{figure}
    \centering
    \includegraphics[width=1\linewidth]{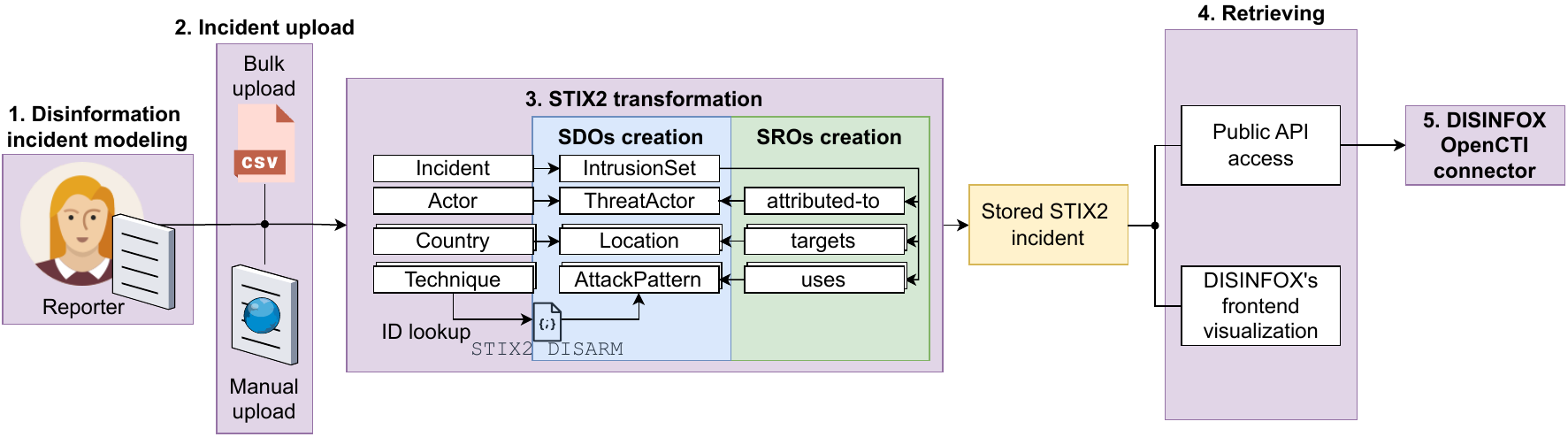}
    \caption{Disinformation incident lifecycle in DISINFOX architecture}
    \label{fig:lifecycle}
\end{figure}

\begin{figure}
    \centering
    \includegraphics[width=\linewidth]{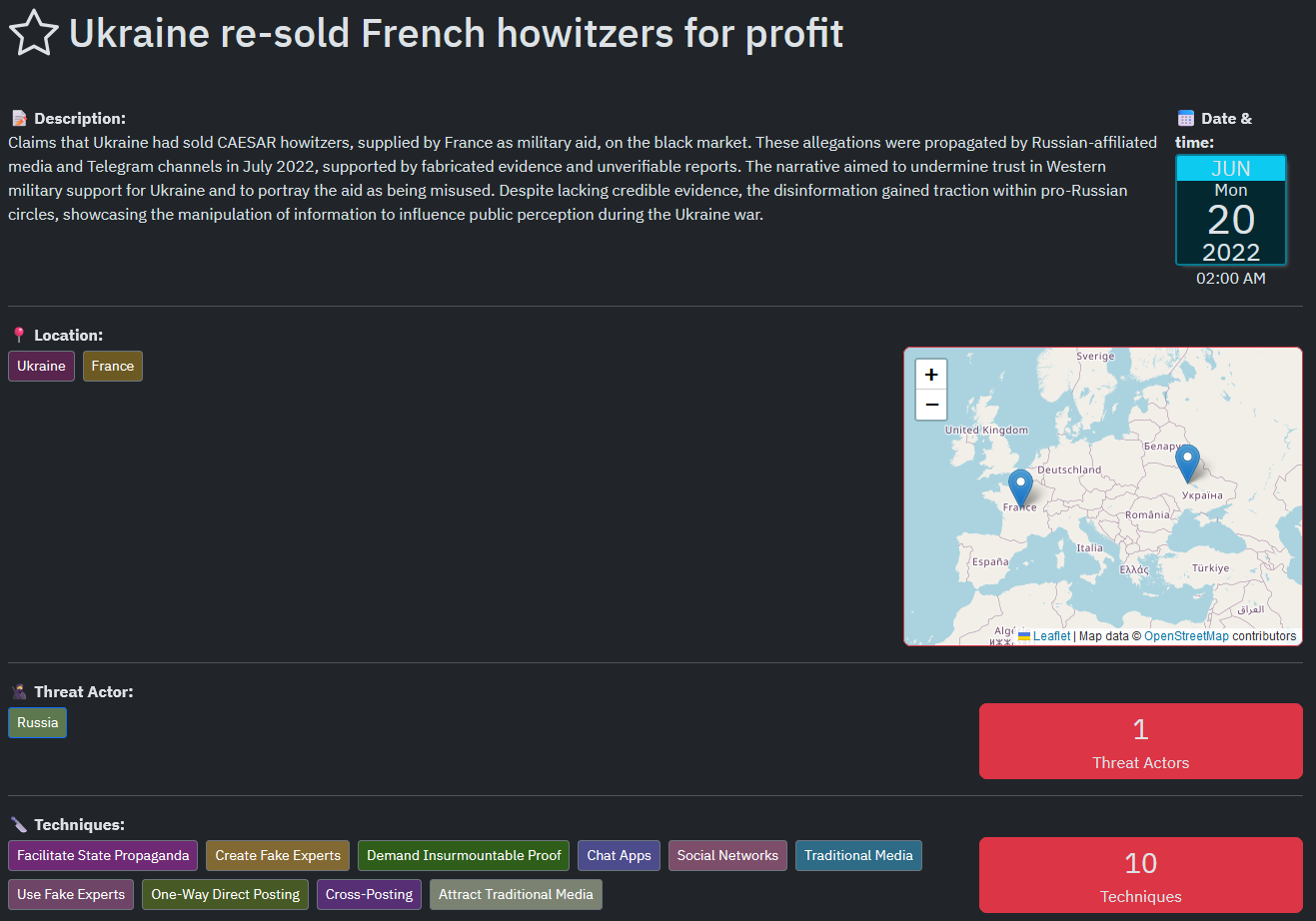}
    \includegraphics[width=\linewidth]{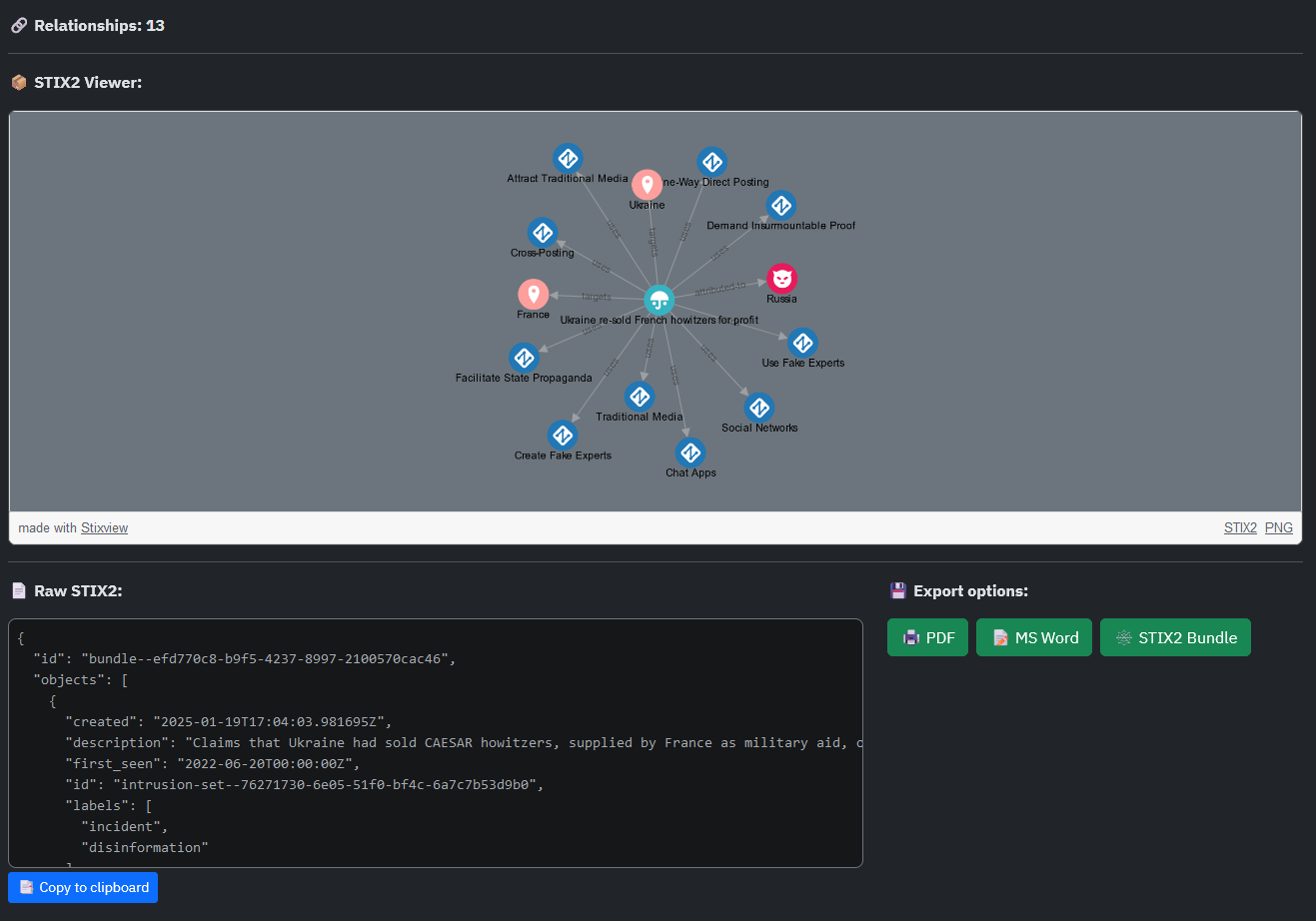}
    \caption{DISINFOX Platform: Frontend visualization of the uploaded URFH incident}
    \label{fig:incident-detail}
\end{figure}

\begin{figure}[h!]
    \centering
    \includegraphics[width=\linewidth]{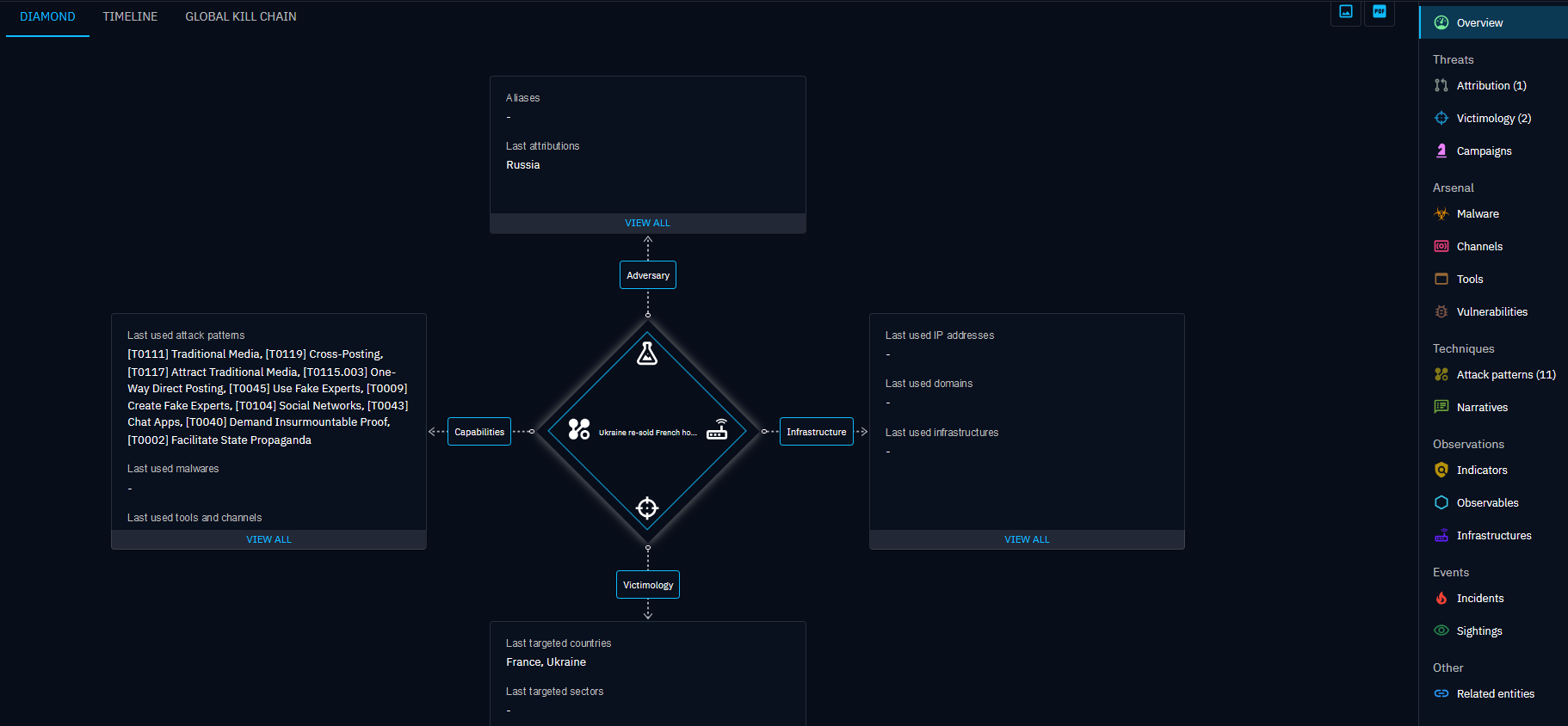}
    \includegraphics[width=\linewidth]{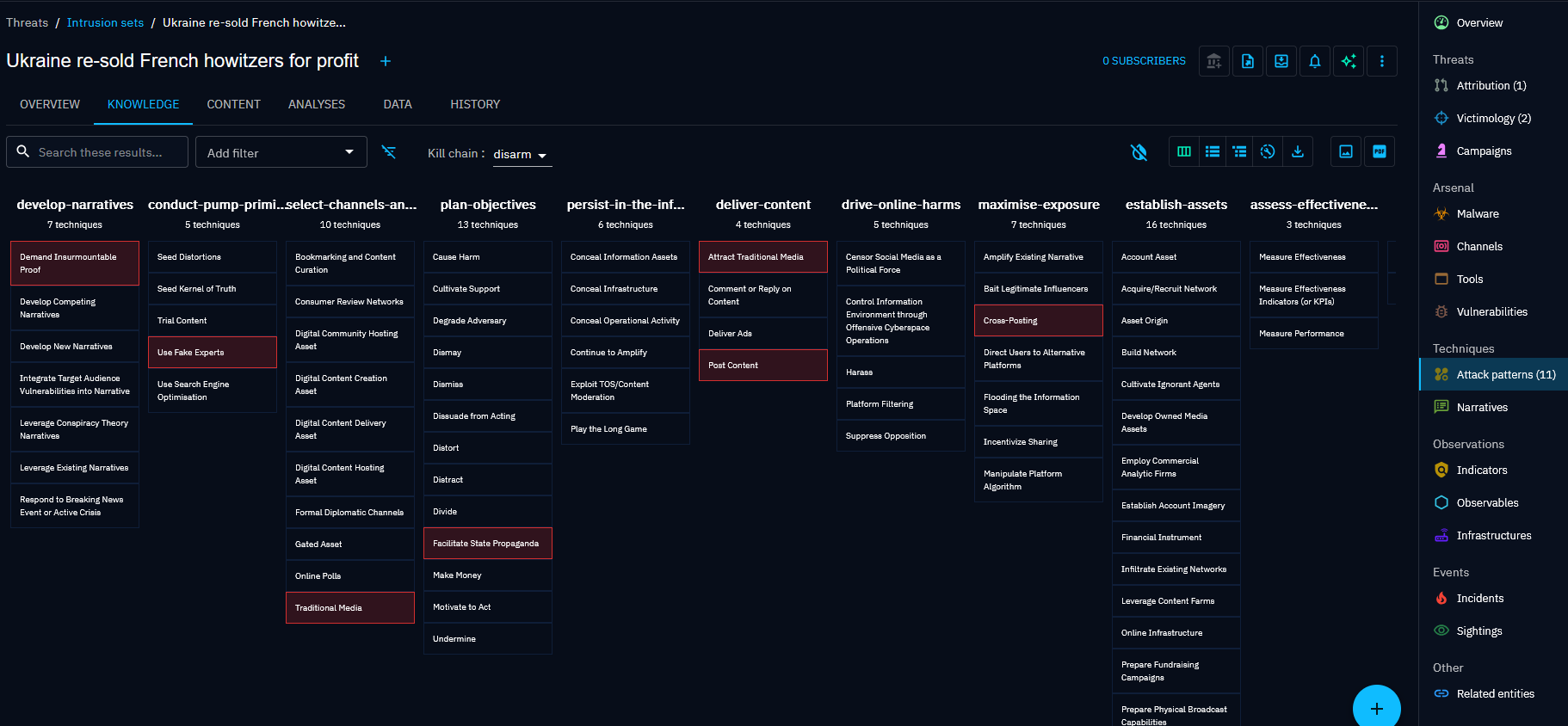}
    \caption{DISINFOX clients: OpenCTI \textit{Knowledge} tab in the consumed URFH incident}
    \label{fig:octi-knowledge}
\end{figure}

\begin{itemize}
    \item \textbf{Frontend visualization}: The web-based interface allows users to browse and interact with stored incidents, presenting relationships between actors, techniques, and regions.  Figure \ref{fig:incident-detail} illustrates the URFH incident on the platform’s web page. The interface provides a detailed view of the incident with maps and knowledge graphs. Users can export the incident as a STIX2 bundle, or Word/PDF files.

    \item \textbf{API access}: Developers and analysts can query incidents via the public API, extracting structured intelligence for automated workflows.
    
    \item \textbf{OpenCTI integration} (Step 5): The custom \DISINFOX OpenCTI connector automatizes ingestion into OpenCTI, where disinformation incidents are visualized. Figure~\ref{fig:octi-knowledge} showcases the OpenCTI \textit{Knowledge} tab of the URFH incident retrieved from DISINFOX by the connector, showing incident's relationships with actors, tactics, and techniques. Additionally, the DISARM matrix is leveraged to categorize techniques, mirroring how MITRE ATT\&CK is used in cyber threat analysis.

\end{itemize}

The lifecycle of disinformation incidents within the \DISINFOX architecture was validated using a dataset of over 100 DISARM-modeled incidents\footnote{https://github.com/CyberDataLab/disinfox/blob/main/backend/data/merged_Foulde_DSRM_additions.csv}. This dataset combines incidents from Margot Fulde-Hardy’s working paper \cite{fulde}, entries from the official DISARM repository\footnote{{https://github.com/DISARMFoundation/DISARMframeworks/blob/main/DISARM_MASTER_DATA/DISARM_DATA_MASTER.xlsx}}, and cases modeled by this work.  

\clearpage

\section{Conclusion and future work}
\label{sec:conclusion}

This work has successfully addressed the challenge of applying CTI methodologies to disinformation incidents. First, through the systematic evaluation of existing frameworks for modeling disinformation, the DISARM framework was the most suitable one. Second, a custom STIX2 representation was defined to capture the DISARM TTPs and facilitate the structured representation of incidents, actors, affected countries, and techniques. This mapping was validated with a real disinformation incident in the context of Russia-Ukraine war. Finally, the CTI-compatible \DISINFOX architecture was developed as an open-source interoperable system designed for serving disinformation incidents to CTI clients. Implemented using a containerized architecture with Docker, the centralized platform stores the existing incidents in STIX2 format, integrating a frontend for user-friendly interaction and a public API for data retrieval. The full incident lifecycle was validated through the implementation of a proof-of-concept \DISINFOX OpenCTI connector, which successfully pushed more than 100 disinformation incidents into OpenCTI clients. Notably, the technological stack used by \DISINFOX architecture (DISARM + STIX2.1 + OpenCTI) aligns with the approach jointly agreed upon by the EU and the US for addressing FIMI, as outlined in the \textit{EU-US Trade and Technology Council’s} fourth ministerial meeting~\cite{ministerialttc}. 


Despite these achievements, certain limitations remain. The dataset currently consists of 118 ingested incidents, which, while sufficient for validation, is relatively small. Expanding the dataset would enhance correlation opportunities and provide deeper insights into disinformation tactics. Additionally, the manual nature of incident modeling using the DISARM framework presents a bottleneck, as analysts must manually label techniques and actors, making large-scale adoption more labor-intensive. Another constraint lies in the STIX2 data model, which, while functional, follows a minimal mapping. Further extending this representation would enrich incident descriptions and improve analytical capabilities. Lastly, the lack of TAXII support in the public API limits standardization in how incidents are shared with external CTI systems, reducing interoperability with platforms that rely on this protocol for structured intelligence exchange.

Future work will address these limitations by focusing on automation and standardization. Expanding the dataset and integrating additional CTI platforms would further validate the system's scalability and practical impact. Additionally, the integration of Large Language Models could significantly reduce the time and expertise required to map disinformation incidents to DISARM TTPs. Finally, aligning \DISINFOX architecture with emerging standardized data models such as DAD-CDM would enhance interoperability and knowledge representation. 

\section*{Acknowledgement}
This study was partially funded by (a) the strategic project ``Development of Professionals and Researchers in Cybersecurity, Cyberdefense and Data Science (CDL-TALENTUM)" from i) the Spanish National Institute of Cybersecurity (INCIBE) and ii) by the Recovery, Transformation and Resilience Plan, Next Generation EU, and (b) by a ``Juan de la Cierva'' Postdoctoral Fellowship (JDC2023-051658-I) funded by the i) Spanish Ministry of Science, Innovation and Universities (MCIU), ii) by the Spanish State Research Agency (AEI/10.13039/501100011033) and iii) by the European Social Fund Plus (FSE+).



\bibliographystyle{elsarticle-num} 
\bibliography{cas-refs}





\end{document}